# Polymerase/nicking enzyme powered dual-template multi-cycled G-triplex machine for HIV-1 determination

Qiuyue Duan,[a] Qi Yan,[a] Yuqi Huang,[a] Wenxiu Zhang,[a] Shuhui Zhao,[a] Gang Yi[*a]

[a]*Key Laboratory of Clinical Laboratory Diagnostics (Ministry of Education of China), Department of Laboratory Medicine, Chongqing Medical University, Chongqing, 400016, P. R. China.*

**Keywords:** DNA nanomachine, HIV, multi-cycled amplification strategy, G-triplex, trace biomarker analysis

## 1. Summary

We proposed a dual-template multi-cycled DNA nanomachine driven by polymerase/nicking enzyme with high efficiency. The reaction system simply consists of two templates (T1, T2) and two enzymes (KF polymerase, Nb.BbvCI). The two templates are similar in structure (X-X'-Y, Y-Y'-C): primer recognition region, primer analogue generation region, output region (3'- 5'), and there's a nicking site between each two regions. Output of T1 is the primer of T2 and G-rich fragment (G3) is designed as the final products. In the presence of HIV-1, numerous of G3 were generated owing to the multi-cycled amplification strategy and formed into G-triplex/ThT complex after the addition of thioflavin T (ThT), which greatly enhanced the fluorescence intensity as signal reporter in the label-free sensing strategy. A dynamic response range of 50 fM-2 nM for HIV-1 gene detection can be achieved through this multi-cycled G-triplex machine, and benefit from the high efficiency amplification strategy, enzymatic reaction can be completed within 45 minutes followed by fluorescence measurement. In addition, analysis of other targets can be achieved by replacing the template sequence. Thus there's a certain application potential for trace biomarker analysis in this strategy.

## 2. Introduction

Increasing number of studies shed light on the significance of monitoring diseases related trace amount biomarkers [1-4] (eg.nucleic acid, protein and small molecule) for the reason that determining the concentration of a molecule precisely can be vital for early diagnosis, treatment, monitoring, and prognosis assessment of a disease[5-8]. As we know, biomarkers are found in low-picomolar concentrations or even femtomolar concentrations in clinical samples [9-11]. Various strategies serve the unmet need for ultrasensitive determination of these biological molecules with great accuracy and specificity. For instance, microarray analysis, PCR (polymerase chain reaction), LAMP (loop-mediated isothermal amplification) reaction and so on [12-15]. However drawbacks such as inconvenient operation procedure, time-consuming and dependant of costly equipment still remain, which leads to limited application and tough development in ordinary laboratory.

*Author for correspondence ( yigang666@cqmu.edu.cn ).

†Present address: Key Laboratory of Clinical Laboratory Diagnostics (Ministry of Education of China), Department of Laboratory Medicine, Chongqing Medical University, Chongqing, 400016, P. R. China.

Several biosensing strategies with promising amplification methods have shown good performance in bioanalysis, especially the emerging of multiple DNA machines with different functions, such as DNA walkers [16], DNA switches [17-18], DNA robots [19] and DNA coppiers [20-21]. Precise and intelligent control of the operation in DNA machines under specific conditions is a necessary condition for itsconstruction. That means, a DNA machine has to respond quickly and powerful in existence of target analyte, while never work with no stimulate. And the amplification efficiency is of great significance for a DNA machine. For better performance, researchers integrated several amplification reactions into one strategy to improve the amplification efficiency [22-25]. However, we found that integration of different amplification methods meets difficulties such as high background caused by non-specific amplification in complex reaction mixture, extended reaction time and troublesome operation on account of complicated thermal cycle. Thus prior DNA machine with ideal amplification efficiency under simple operation is in great demand.

Polymerase/nicking endonuclease fueled isothermal exponential amplification with high efficiency has attached researchers' interests as an ideal candidate to construct various DNA machine-based analytical strategies. DNA machines driven by specific polymerase and endonuclease run programmatically and accurately owing to precise paring of four bases in double helix and high efficient extending or nicking function of enzymes [26-28]. After specific recognition of target analyte and template strand or recognition probe, enzymes in these methods act like "engine" of the machine, driving the machine in high speed automatically [28-29].

Benefiting from the remarkable specificity of Watson-Crick base pairing rule, nucleic acid played a vital role in various machine-like amplification strategies for its programmable sequence [30-33]. Recent reports found stable DNA secondary structure named G-triplex by shortening the length of G-quadruplex sequence to a 13-mer sequence folding into three G-tracts with structure similar to G-quadruplex and conducts the same biofunction such as catalysis and light-up fluorescence [33-34]. Overcoming the limited modulation of G4 caused by the relatively longer 22-mer sequence, the G-triplex is easier to be generated and freed off from the template. And G-triplex template based polymerization/nicking reaction system brings higher signal amplification efficiency for faster generation speed of G-triplex[35]. Moreover, combining with ThT, the G-triplex increases fluorescence intensity even more dramatically than G-quadruplex [34]. Due to inherent advantages such as cheap, stable, and easy to synthesize of nucleic acid, this non-Watson-Crick DNA secondary structure with 3 G-tracts has been widely used in application in development of biosensing platforms [35-36].

Inspired by the above work, a simple dual-template multi-cycled G-triplex machine based on polymerase/nicking enzyme fueled isothermal exponential amplification for ultrasensitive bioanalysis was constructed. The DNA machine simply consisted of two templates (T1, T2) and two enzymes (KF ploymerase, Nb.BbvCI), and G-triplex/ThT complex was used as signal reporter. Considering the infection of HIV worldwide, high mortality of acquired immunodeficiency syndrome (AIDS) and increasing requirements for early diagnosis, therapy and prevention of virus's propagation [37-38], HIV-1 gene was chosen as target analyte. With HIV-1 binding to T1, amounts of output strands Y produced in extending reaction of T1 stared up new reaction cycle of T1 as feedback amplification and series of downstream polymerization/nicking reaction of T2 automatically. As a result, massive G3 were produced through this multi-cycled G-triplex machine and formed into G-triplex/ThT complex with the existence of the inexpensive water-soluble fluorogenic dye ThT [39]. Owing to these merits, a new one-pot and label-free G-triplex machine for fast and ultrasensitive determination of HIV-1 with low background was constructed.

# 3. Materials and Methods

## 3.1. Reagents and materials

HPLC purified oligonucleotides used in this research (As listed in Table S1†) were supplied by Sangon Biotech Inc. (Shanghai, China), and fully dissolved in the ice box with TE buffer (10 mM Tris HCl, 1 mM EDTA, pH 8.0) into a storage concentration of 10 μM. 1 × TE buffer, 10 mM deoxynucleotide triphosphates (dNTPs), 6 × DNA loading buffer, 30% acrylamide/bis solution, ammonium persulfate (APS) were obtained from Sangon Biotechnology Co. Ltd (Shanghai, China). N,N,N′,N′-tetramethylethylenediamine (TEMED) and 20 bp DNA Ladder (Dye Plus) used for polyacrylamide gel electrophoresis was acquired from TaKaRa Biotech (Dalian, China). GoldView was acquired from Solarbio LIFE SCIENCES (Beijing, China). Thioflavin T (ThT) was purchased from BBI LIFE SCIENCES CORPERATION (Shanghai, China) and resolved with ultrapure water into 100 μM for fluorescent measurement. Nb.BbvCI endonuclease and Klenow fragment (3′-5′ exo-) polymerase (KF polymerase) used in this study were served by New England Biolabs (Beijing, China). Ultrapure water used for solution preparing was obtained from a Millipore water purification system with a resistivity of 18.2 MΩ cm. Human serum samples for recovery experiment were offered by the First Affiliated Hospital of Chongqing Medical University.

## 3.2. Apparatus and instruments

A Cary Eclipse Fluorescence Spectrophotometer (Agilent Technologies, Palo Alto, CA) was used for all fluorescence spectra measurements using a quartz fluorescence cuvette (optical path length of 1.0 cm) at an excitation wavelength of 442 nm. And 10 nm was set as both excitation and emission slit widths. The DYY-6C electrophoresis analyzer (Liuyi Instrument Company, China) and Bio-Rad ChemDoc XRS (Bio-Rad, USA) were used for polyacrylamide gel electrophoresis.

## 3.3. Fluorescence spectrophotometer analysis of HIV-1 gene

A conventional operation of the multi-cycled G-triplex machine was performed in a one-step way as described below. To reduce the background signal, the reaction liquid was divided into two parts named mixture A and mixture B. Mixture A included various concentrations of HIV-1 gene, two templates (T1 and T2), dNTPs and NEB buffer 2. Mixture B contained KF polymerase, Nb.BbvCI and NEB buffer 2. And the two parts were fully mixed in an ice box before a 37 °C incubation for 45 min at a final volume of 10 μl. After that, the mixture was heated at 85 °C for 10 min to inactivate the enzymes. Finally, 5 μM ThT, 50 mM KCl and ultrapure water were added in to make a 50 μl volume solution for fluorescence measurement.

## 3.4. Gel Electrophoresis analysis

The multi-cycled G-triplex machine was analyzed by 12% native polyacrylamide gel electrophoresis (nPAGE) in 1 × TBE buffer at a constant voltage of 120 V for 45 min. GoldView was used for gel staining subsequently, and after an exposure to the ultraviolet light, the dyed gel was visualized via a gel imaging system (Bio-Rad Laboratories, USA).

## 3.5. Circular dichroism (CD) measurements

Circular dichroism (CD) measurements were carried out on the Chirascan CD spectrometer (Applied Photophysics Ltd., UK) in a 1 mm path length quartz cuvette. Three scans were performed for each sample at room temperature under the following parameters: range 200–500 nm, bandwidth 1 nm, scanning speed 200 nm/min and a response time of 0.5 s [34].

# 4. Results and discussion

## 4.1. Principle of the dual-template multi-cycled G-triplex machine

Principle of the G-triplex machine is illustrated in scheme 1. The one-pot label-free G-triplex machine was formed by polymerase/nicking enzyme fueled multi-cycled exponential amplification. The two templates named T1, T2 were similar in structure (X-X'-Y, Y-Y'-C): primer recognition region, primer analogue generation region, output region (3'-5'), and there's a nicking site between every two regions. With the existence of HIV-1 gene, HIV-1 gene hybridized with T1 and switched on the polymerase/nicking endonuclease powered DNA machine automatically. In brief, the target analyte HIV-1 DNA triggered cycle 1 with generation of many X' and Y. Cycle 2 was induced by plenty of X' which led to more production of Y subsequently. Considerable amounts of Y started up cycle 3 as primer of T2 and brought massive amounts of Y' and G3. Finally, G3 emerged in large number owning to the feedback amplification conducted by Y' in cycle 4. Greatly enhanced fluorescence intensity could be obtained with numerous G3 forming into G-triplex/ThT complex.

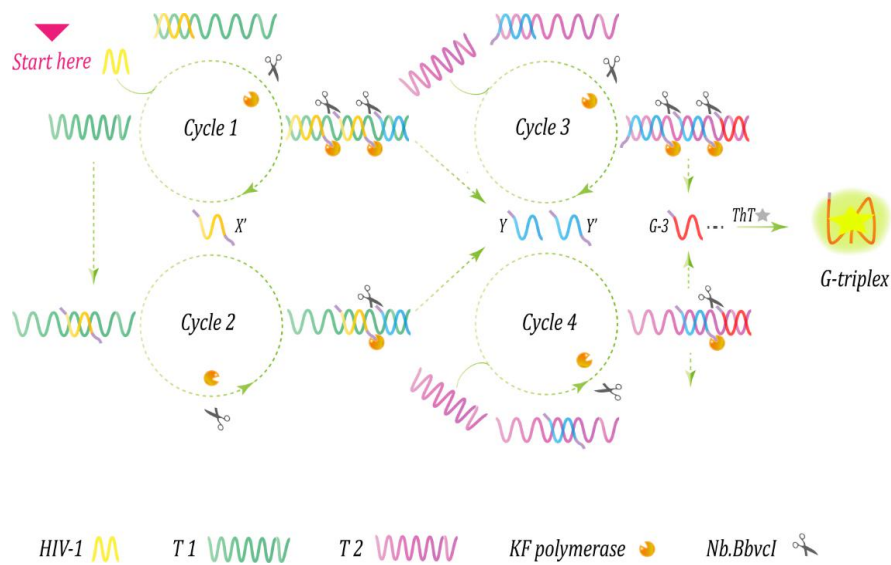

**Scheme 1. Principle of the dual-template multi-cycled G-triplex machine for HIV-1 determination.**

## 4.2. Characterization of G-triplex

CD and fluorescence spectrophotometer analysis was used for characterization about the structure and biofunction of G-triplex. As depicted in Figure 1A, there're a positive peak at 265 nm and a negative peak at 240 nm in spectrum of G-triplex, which match the typical parallel stand arrangement, suggesting formation of parallel G-triplex [34]. And this can be well distinguished by the characteristic of double-stranded DNA and single-stranded DNA which possess a positive peak at 280 nm.[40] The two peaks remained unchanged after the addition of ThT, and a new negative peak appeared in the CD spectrum at 425 nm, indicating an intercalation mode of binding without change of parallel structure in the G-triplex/ThT complex. Figure 1B illustrated that fluorescence signal was extremely low in the presence of either ThT or G-triplex, while greatly enhanced fluorescence intensity could be measured in G-triplex/ThT complex. These results are good proof of the formation of G-triplex and its biological function in enhancing fluorescence intensity of ThT.

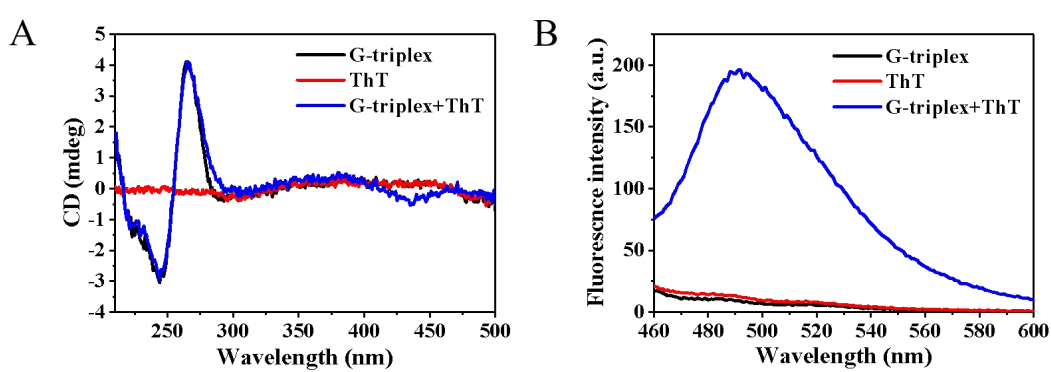

**Figure 1. Characterization of G-triplex. (A) CD spectra of 5 μM G-triplex, 100 μM ThT and the mixture of G-triplex and ThT. (B) Fluorescence spectrophotometer analysis of 1 μM G-triplex, 5 μM ThT, and the mixture of G-triplex and ThT (25 mM Tris–HCl buffer (pH 7.4) containing 50 mM KCl).**

## 4.3. Feasibility of the G-triplex machine

To evaluate feasibility of the G-triplex machine, fluorescence spectrum analysis and nPAGE was carried on. Figure 2A was the diagram of nPAGE. Lane 1, Lane 2 and Lane 3 were HIV-1 DNA, T1 and T2. Lane 7 and Lane 8 were final products G3 and 20 bp DNA marker. We can see products band in lane 5 was the same as the G3 band in lane 7, this manifested a successful construction of the amplification strategy. The G3 band didn't appear in lane 4 or lane 6 for absence of Nb.BbvCI in lane 4, and lack of HIV-1 DNA in lane 6, thus the polymerization/nicking reaction can hardly carry out. The results certificated that the multi-cycled amplification strategy has been established successfully.

In order to further prove the construction of this G-triplex machine, fluorescence spectrum analysis was putted on. In addition, we constructed a traditional two-cycled strategy with a template named S1, and conducted a fluorescence spectrum analysis of this two-cycled strategy compared with our multi-cycled strategy towards 1 nM HIV-1 DNA. As depicted in Figure 2B, there's only weak fluorescence in blank, and a high fluorescence intensity was detected in our multi-cycled strategy with a signal-to-noise ratio (SNR) up to nearly 11 times, see the red curve. Moreover, fluorescence intensity in the two-cycled strategy was much lower than that in our strategy, see the pink curve. These results revealed that construction of the dual-template multi-cycled G-triplex machine was successful and it owned admirable performance.

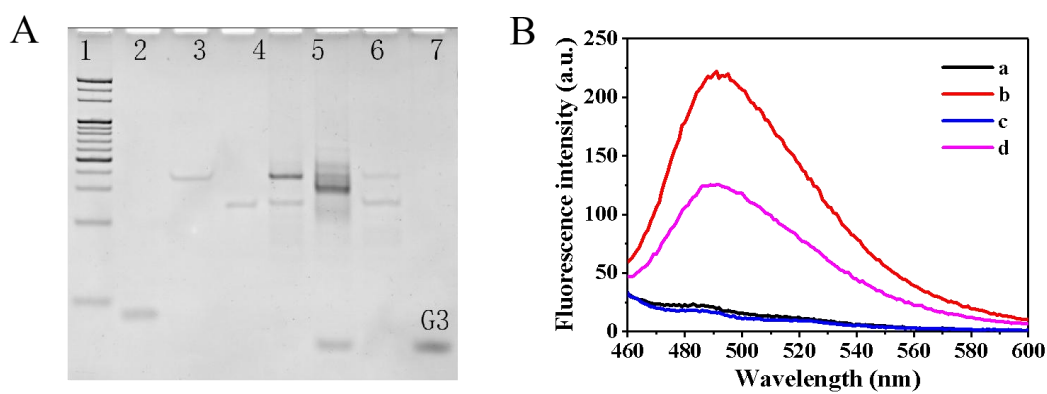

**Figure 2.** Feasibility of the dual-template multi-cycled G-triplex machine. (A) Verification of the dual-template multi-cycled amplification strategy through the 12% nPAGE. Line 1, HIV-1; Line 2, T1; Line 3, T2; Line 4, T1+T2+HIV-1+KF; Line 5, T1+T2+HIV-1+KF+Nb.BbvCI; Line 6, T1+T2+KF+Nb.BbvCI; Line 7, G3; Line 8, 20 bp DNA marker (Sample concentrations are 2 μM in Line 1, Line 2, Line 3 and 3 μM in Line 7. Concentration of T2 and T1 are 1 μM and 100 nM separately in Line 4, 5 and 6. (B) Fluorescence spectroscopy analysis of the dual-template multi-cycled G-triplex machine compared to two-cycled amplification strategy. Curve a, blank of the multi-cycled amplification strategy; Curve b, signal of the multi-cycled amplification strategy towards 1 nM HIV-1 DNA; Curve c, blank of two-cycled amplification strategy, Curve d, signal of two-cycled amplification strategy towards 1 nM HIV-1 DNA.

## 4.4. Optimization of experimental conditions

Exploration of optimal reaction conditions including molar ratio between T1 and T2 (T1/T2), the reaction time, dosage of KF polymerase and Nb.BbvCI were conducted for acquisition of optimal performance. As we know, many polymerase/nicking reaction system meets the high background signal caused by non-specific amplification of excessive templates, and we found that inappropriate molar ratio of templates led to high background signal as well. Thus, we gave priority to optimization of T1/T2. A series of T1/T2 was set under a total concentration of 100 nM to get the most powerful DNA machine. As shown in Figure 3A, the SNR has been greatly improved with T1/T2 reduced from 1/10 to 1/160, and reached the highest level under a molar ratio of 1/80. When T1/T2 decreased to 1/160, the SNR dropped a little, this might be associate with a slightly decline of the signal amplification efficiency due to fewer T1 in the reaction system. Therefore, 1/80 was set as the most suitable T1/T2 for subsequent experiments.

We next investigated the enzymatic reaction time. As shown in Figure 3B, extension of the reaction time improved the SNR in the beginning, and the ideal SNR appeared at the reaction time of 45 minutes. As the reaction time prolonged, the SNR dropped dramatically. Thus, 45min was adopted for the multi-cycled amplification reaction time accordingly.

KF polymerase is the energy provider of this DNA machine. We analyzed performance of the sensor at different dosage of KF polymerase subsequently. As illustrated in Figure 4A, with the dosage of KF polymerase increased to 0.15 U, the fluorescence signal enhanced significantly and the background remained low. Further increasing of the KF polymerase heightened the background rapidly and decreased the SNR as presented in Figure 4B. The results indicated that working efficiency of the G-triplex machine had reached the best with dosage of KF polymerase at 0.15 U, so we determined the satisfied dosage of KF polymerase was 0.15 U.

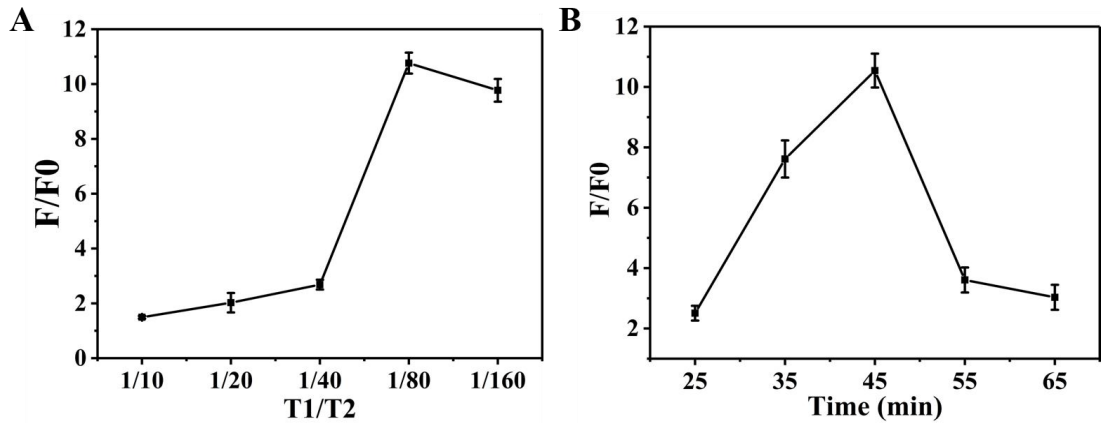

**Figure 3.** Optimization of (A) T1/T2. (B) the reaction time (1 nM HIV-1, T1/T2 at 1/80, 0.15 U KF, 5 U Nb.BbvCI). Error bars indicates standard deviation of tests for three times.

Finally, another part of engine in this G-triplex machine, dosage of the Nb.BbvCI was explored. In Figure 4C, fluorescenceintensity improved as the dosage of Nb.BbvCI increased, and reached the highest point at 5 U. With Nb.BbvCI dosage rose to 6 U, fluorescence intensity dropped a bit. This might due to oversaturation of the Nb.BbvCI. We can see that the SNR elevated fast in view of the increasing energy provided by Nb.BbvCI, and then leveled off at 5 U as shown in Figure 4D accordingly. Therefore, 5 U was considered as the optimal Nb.BbvCI dosage to ensure well performance of the sensor.

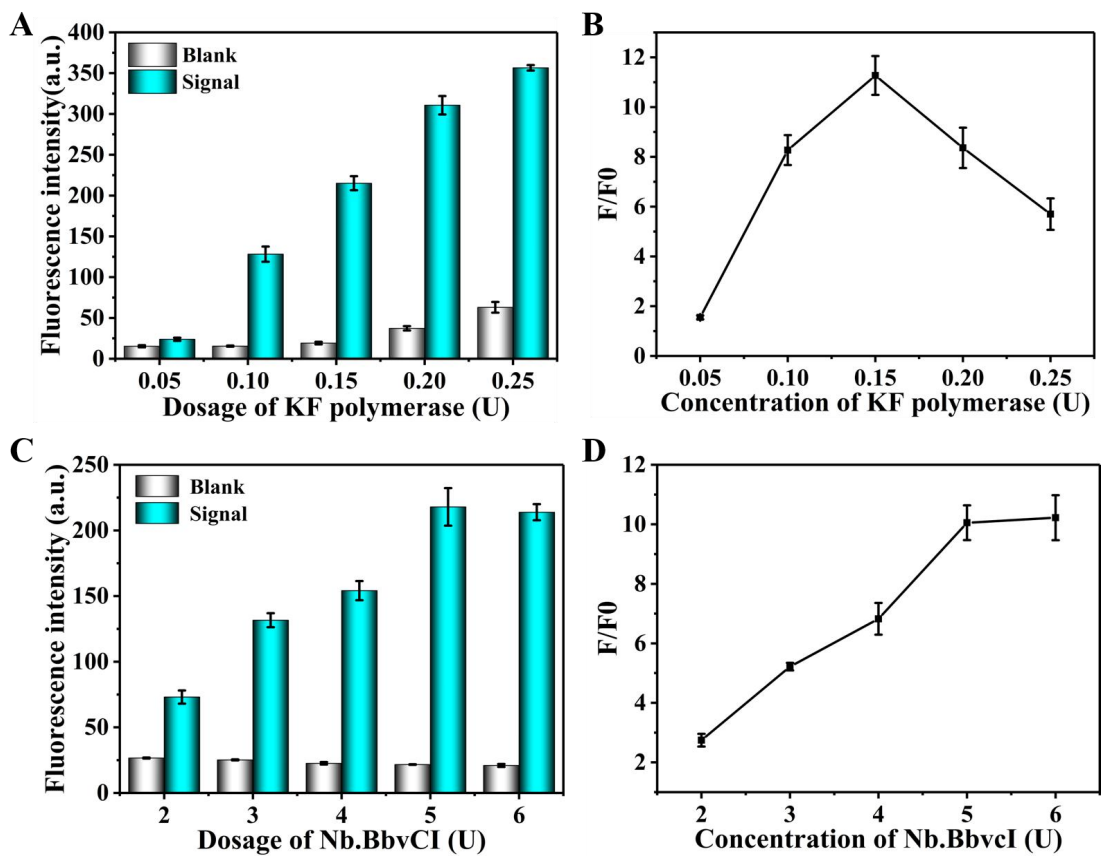

**Figure 4.** Optimization experiment. (A) Effects of different dosage of KF polymerase on the fluorescence intensity (1 nM HIV-1, T1/T2 at 1/80). (B) SNR towards different dosage of KF polymerase. (C) Influence of the dosage of Nb.BbvCI (1 nM HIV-1, T1/T2 at 1/80, 0.15 U KF). (D) SNR towards different dosage of Nb.BbvCI. Error bars indicates standard deviation of tests for three times.

## 4.5. Analytical performance of the biosensing strategy

Sensitivity, selectivity and stability are primarily included in analytical performance of a sensor. Herein, optimal conditions were used for exploring of the analytical performance. First of all, different concentrations of HIV-1 gene (50 fM-2 nM) were used for fluorescence intensity measurements in evaluation of the sensitivity. As shown in Figure 5A, higher fluorescenceintensity was measured with target gene increased gradually. This can be attributed to the fact that more HIV-1 DNA produced more intermediate strand Y and induced massive production of G3, which greatly enhanced the fluorescence signal of ThT. And there's a linear relationship between logarithm of the HIV-1 gene concentration (C) and the corresponding fluorescence intensity (F) in an HIV-1 DNA concentration range of 50 fM-2 nM as described in Figure 5B. Corresponding regression equation is F = 41.20 lg C + 209.50 with a correlation coefficient of 0.9971. According to the 3 σ rule, the limit of detection (LOD) was calculated to be 30.95 fM. And this has been a prior analysis ability compared with many reported sensing strategies (Table S2†). Good performance of this detection strategy benefited from the superior magnification effect of the multi-cycled amplification strategy in this constructed DNA machine and the high efficiency of the cooperative enzyme-driven reaction.

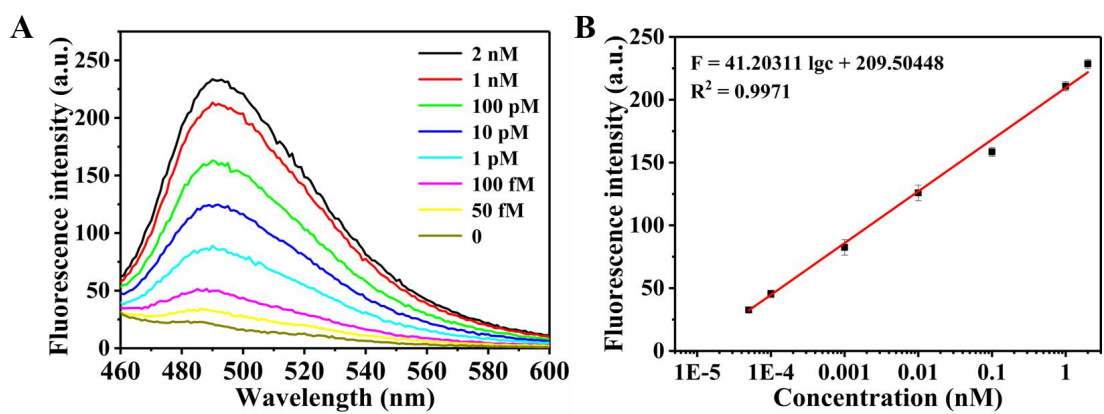

Figure 5. Sensitivity of the multi-cycled G-triplex machine. (A) Fluorescence spectrum of the dual-template multi-cycled DNA machine with target HIV in different amounts: 0, 50 fM, 100 fM, 1 pM, 10 pM, 100 pM, 1 nM, 2 nM. (B) Plot the linear relationship between the fluorescence intensity and the concentration of HIV-1 from 50 fM to 2 nM. Error bars indicates standard deviation of tests for three times.

To further study the selectivity of this intelligent DNA machine, a non-complementary nucleic acid and three mismatched strands (single-base mismatched, two-base mismatched, and three-base mismatched DNA) named rDNA, DNA1, DNA2 and DNA3 were used for fluorescence measurements in the same concentration. Figure 6A illustrated fluorescence response of these nucleic acid and HIV-1 in the same assay condition. The rDNA led to extremely low fluorescence response and signal response of DNA1, DNA2, DNA3 were much lower than the significant fluorescence intensity rose in HIV-1 gene as well. This clearly revealed that the developed DNA machine possessed high specificity in target DNA identifying.

Finally, we analyzed the stability of the sensing strategy by placing the sample at 4°C for 7 days. During the seven-day storage period, the peak intensity of the fluorescence spectrum did not change significantly. As shown in Figure 6B, the RSD value of the seven-day test period was 3.96%, indicating a superior stability of this multi-cycled G-triplex machine during storage.

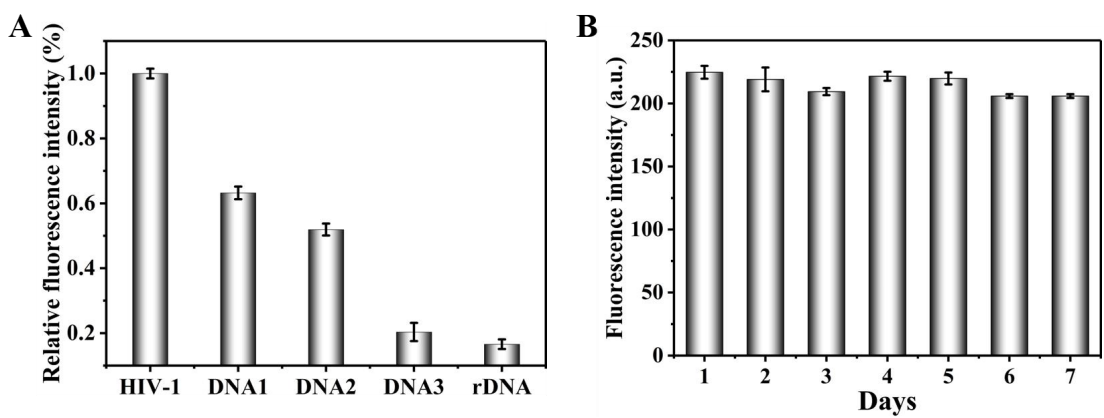

Figure 6. (A) Specificity of the dual-template multi-cycled G-triplex machine in analysis of HIV-1. (B) Stability of the constructed G-triplex machine in a 7-day storage under 4 ℃. Error bars indicates standard deviation of tests for three times.

### 4.6. Interference studies

Considering the complicated composition of actual sample, recovery experiment was carried out in 20 times diluted serum for evaluation of the utilization potential. Different concentrations (100 fM, 10 pM, 1 nM) of HIV-1 gene were added to the diluted serum, and corresponding fluorescence intensity were measured then. As demonstrated in Table 1, recovery rates of the samples ranged from 96.42% to 104.93% with RSD of 0.84%, 1.86% and 1.79% separately. The above data showed that interference of the actual sample was insignificant, revealing the powerful applicability and certain application potential for clinical diagnostics of this method.

**Table 1. Recovery test of target DNA in human serum diluted for 20 times.**

| Spiked sample concentration | Detected concentration | Recovery | RSD |
| --- | --- | --- | --- |
| 100 fM | 104.93 fM | 104.93% | 0.84% |
| 10 pM | 9.64 pM | 96.42% | 1.86% |
| 1 nM | 1.02 nM | 102.15% | 1.79% |

## 5. Conclusion

We have proposed a new one-pot and label-free G-triplex machine for fast and ultrasensitive determination of HIV-1 with low background. The constructed sensing strategy introduced multiple amplification cycle and shortened the reaction time so that the reaction can be completed within 45 minutes. Moreover, this simple reaction system didn't need complicated temperature cycle, troublesome operation or costly equipment. It performed in a one-step way in simple reaction system (two templates and two enzymes), thus decreasing non-specific amplification and maintaining a low background. At the same time, the established polymerase/nicking endonuclease driven G-triplex machine possesses higher efficiency ensuring excellent sensitivity, and realizing an ultrasensitive detection of HIV-1 DNA as low as 30.95 fM with a wide linear response range of 50 fM-2 nM. In addition, utilization of the G-triplex/ThT complex as a fluorescent probe didn't require extra modification of nucleic acids, thereby reducing the cost. Furthermore, by replacing sequence of the target recognition region in template, detection of any other objects can be achieved as well. This strategy provides a new idea for the detection of trace substances.

**Funding Statement**

This work was supported by the Science and Technology Research Program of Chongqing Yuzhong District Science and Technology Commission (Grant No. 20180127) and the Special Fund Project Key Laboratory of Clinical Laboratory Diagnostics (Ministry of Education).

**Competing Interests**

We declare we have no competing interests.

# Supporting Information

**Polymerase/nicking enzyme powered dual-template multi-cycled G-triplex machine for HIV-1 determination.**

Qiuyue Duan,[a] Qi Yan,[a] Yuqi Huang,[a] Wenxiu Zhang,[a] Shuhui Zhao,[a] Gang Yi[*a]

[a] *Key Laboratory of Clinical Laboratory Diagnostics (Ministry of Education), College of Laboratory Medicine, Chongqing Medical University, Chongqing 400016, China*

---

\* Corresponding author. Tel: +86 23 68485240; Fax: +86 23 68485239; E-mail address: yigang666@cqmu.edu.cn (G. Yi).

**Table S1. Oligonucleotides used in this work[a].**

| Oligonucleotides | Sequence (from 5′ to 3′) |
| --- | --- |
| T1 | TTGT GCA A ATA ACCCCTCAGCAG TGT GGA AAA TCT CTA GCCCTCAGCAGT CAG TGT GGA AAA TCT CTA GC |
| T2 | CCC GCC CTA CCC ACCTCAGCTTGT GCA A ATCCTCAGCTTGT GCA A ATA ACCCCTCA |
| S1 | CCC GCC CTA CCC ACCTCAGCTTGT GCA A ATCCTCAGCAGT CAG TGT GGA AAA TCT CTA GC |
| G3 | TGA GGT GGG TAG GGC GGG |
| HIV-1 | GCTAGAGATTTTCCACACTGACT |
| DNA1 | GCAAGAGATTTTCCACACTGACT |
| DNA2 | GCTAGAGATTGTCCACACTCACT |
| DNA3 | GCAAGAGATTGTCCACACTCACT |
| rDNA | TGGCCCTATACGATCATGGATCA |

[a]T1, T2: template of this multi-cycled G-triplex machine. S1: template of the two-cycled amplification strategy. The highlighted region in orange is the target recognition region, and the nicking site of KF polymerase is highlighted in blue. The region highlighted in red represents the complementary region of G3. G3: products of T2 that could fold into a G-triplex. DNA1, DNA2 and DNA3: one-base, two-base, and three-base mismatched HIV-1 DNA. rDNA: non-complementary strand.

**Table S2. Comparison of the reported methods in nucleic acid detection.**

| Detection method | Strategy | Dynamic range | Limit of detection | Detection time | Reference |
|---|---|---|---|---|---|
| Fluorescence detection | TSDR based cascade | 10 pM - 10 nM | 3.2 pM | 240min | [41] |
| Colorimetric detection | DFA machine | 0.01 nM - 150 nM | 10 pM | 120 min | [42] |
| Fluorescence detection | PMB based SDA | 10 pM - 200 nM | 10 pM | 90 min | [43] |
| Fluorescence detection | hairpin/DNA ring ternary probe based RCA | 10 fM - 10 nM | 1.51 fM | 220min | [44] |
| Fluorescence detection | N-RCA and SSDA based casecade amplification | 5 pM - 20 nM | 76 fM | 180min | [45] |
| Fluorescence detection | RCA-CDM and CSDA-BDM basded dual DNA machine | 1 pM - 50 nM | 10 pM | 150 min | [23] |
| Fluorescence detection | TRA and T-SDA based table-tennis-motion-type DNA machine | 0.5 pM - 500 nM | 10 pM | 120 min | [46] |
| Fluorescence detection | Dual-template multi-cycled G-triplex machine | 50 fM - 2 nM | 30.95 fM | 45 min | This method |